\documentclass[twocolumn,]{aastex63}

\usepackage{amsmath}
\usepackage[ruled,vlined,]{algorithm2e}
\usepackage{lineno}
\usepackage{booktabs}
\definecolor{red}{HTML}{c1161b}

\shorttitle{Optimal Weighting}
\shortauthors{Peirson et al.}
\graphicspath{{./}{figures/}}

\begin{document}

\title{Towards Optimal Signal Extraction for Imaging X-ray Polarimetry}

\author[0000-0001-6292-1911]{A. L. Peirson}
\affiliation{Kavli Institute for Particle Astrophysics and Cosmology, Dept. of Physics \\
Stanford University \\
Stanford, CA, 94305}

\author[0000-0001-6711-3286]{Roger W. Romani}
\affiliation{Kavli Institute for Particle Astrophysics and Cosmology, Dept. of Physics \\
Stanford University \\
Stanford, CA, 94305}










\begin{abstract}
We describe an optimal signal extraction process for imaging X-ray polarimetry using an ensemble of deep neural networks. The initial photo-electron angle, used to recover the polarization, has errors following a von Mises distribution. This is complicated by events converting outside of the fiducial gas volume, whose tracks have little polarization sensitivity. We train a deep ensemble of convolutional neural networks to select against these events and to measure event angles and errors for the desired 
gas conversion tracks. We show how the expected modulation amplitude from each event gives an optimal weighting to maximize signal-to-noise ratio of the recovered polarization. Applying this weighted maximum likelihood event analysis yields sensitivity (MDP$_{99}$) improvements of $\sim$10\% over earlier heuristic weighting schemes and mitigates the need to adjust said weighting for the source spectrum.  We apply our new technique to a selection of astrophysical spectra, including complex extreme examples, and compare the polarization recovery to the current state of the art.

\end{abstract}


\section{Introduction} \label{sec:intro}
Imaging polarimetry in the classical soft X-ray band (1-10\,keV) offers many novel probes of compact objects and non-thermal X-ray nebulae. The Imaging X-ray Polarimetric Explorer (IXPE) \citep{weisskopf_overview_2018,odell_imaging_2018}, planned for launch Nov. 2021, will provide our first polarization images of a variety of X-ray source classes. IXPE’s sensitivity is limited by the track analysis algorithm used to recover source polarization, spatial structure and energy, given a measured set of photo-electron track images. In \citet[][hereafter P21]{peirson_deep_2021} we studied a set of simulated IXPE Gas Pixel Detector (GPD) events and demonstrated how convolutional neural network (CNN) track measurement can provide substantial improvements to the current state of the art in polarization recovery. In this work we improve on these results while simplifying the overall analysis pipeline from track reconstruction to polarization prediction. Further, we show that our analysis is near-optimal. We demonstrate this analysis on several simulated astrophysical spectra: generic power-laws, an intermediate synchrotron peak (ISP) blazar 
and a bright accreting X-ray binary, comparing performance to IXPE's default moment analysis scheme \citep{bellazzini_novel_2003}. This range of spectra illustrates how our revised weighting performs well over a broad range of models, eliminating the need to recalibrate weights for each source spectrum. While the results shown here are specific to IXPE’s GPDs, the methods are general, and can be applied to other imaging detector geometries.

In the $1-10$keV range the cross-section for photoelectron emission is proportional to cos$^2(\theta - \theta_0)$, where $\theta_0$ is the normal incidence X-ray's electric vector position angle (EVPA) and $\theta$ the azimuthal emission direction of the photoelectron. Specifically, the emission angles $\theta$ follow the distribution

\begin{equation}
    \theta \sim \frac{1}{2\pi} \big(1 + p_0\mu_{100}\cos[2(\theta - \theta_0)] \big),
    \label{eqn:prob}
\end{equation}

where $0 \leq p_0 \leq 1$ is the true polarization fraction of the source, $-\pi/2 \leq \theta_0 < \pi/2$ is the true source EVPA (the expected direction where the $\theta$-distribution peaks), and $\mu_{100}$ is the modulation factor. The modulation factor, determined by the interaction physics and polarimeter properties, is defined as the amplitude of the azimuthal modulation measured for a 100\% polarized source (i.e. for $p_0 = 1$), so $0 \leq \mu_{100} \leq 1$. The modulation factor $\mu_{100}$ is a strong function of photon energy and of the track reconstruction algorithm. By measuring a data set of individual photoelectron emission angles $\{\theta_i\}_{i=1}^N$, one can recover the above distribution to extract the source polarization parameters ($p_0$,\,$\theta_0$). For imaging X-ray polarimeters one also wishes to reconstruct the X-ray absorption (conversion) points and photon energy.

The sensitivity of a track reconstruction algorithm 
is typically measured by its minimum detectable polarization (MDP$_{99}$) defined as:
\begin{equation}
    {\rm MDP_{99}} \approx \frac{4.29}{\mu_{100}\sqrt{N}}~,
    \label{eqn:MDP}
\end{equation}
where $N$ represents the total number of observed events. This is the 99\% confidence upper limit on polarization fraction $p_0$ for a 0\% polarized source. A lower MDP$_{99}$ is better. It is equivalent to the reciprocal of the signal-to-noise ratio, where we have approximately Poisson counting noise $1/\sqrt{N}$ and the recovered signal is represented by $\mu_{100}$. In an event-weighted analysis $N \rightarrow N_{\rm eff} < N_{\rm events}$. 
Therefore, the best track reconstruction algorithm will maximize $\mu_{100}\sqrt{N_{\rm eff}}$.

The default track reconstruction method for the GPD is a moment analysis described by \citet{bellazzini_novel_2003}. Impressive accuracies for the absorption point and EVPA angle are achieved from a simple weighted combination of track moments. Track energy estimates are proportional to the total collected GPD charge. The track ellipticity also provides a rough proxy for track reconstruction quality. High ellipticity tracks typically have more accurate angle estimates.

We showed in P21 that an ensemble of CNN trained on simulated photoelectron tracks can be used to predict not only point estimates of photoelectron angles $\{\hat{\theta}_i\}$, but also the total Gaussian uncertainty (standard deviation) on these estimates $\{\hat{\sigma}_i\}$. We used these uncertainty estimates as weights in a maximum likelihood estimation to predict source polarization parameters $(p_0, \theta_0)$, improving on moments-analysis accuracy. Our NN approach also significantly decreased photon absorption point and energy resolution errors, estimating these simultaneously with $\hat{\theta}_i, \hat{\sigma}_i$.   

Although our previous approach yielded improved polarimetry results, the weights were heuristic and some aspects complicated practical applications. For example, our event weighting scheme required tuning for individual spectra through a power-law weighting with index $\lambda$, designed to approximate the angle error distribution for a given source spectrum. Thus the weights were sub-optimal, especially over large spectral bands.

Here we adopt a more general approach, studying a set of simulated IXPE event images to define event weights which give improved sensitivity and more uniform applicability. 
To improve event weight fidelity we incorporate an event classification scheme which helps separate good GPD events, converting in the gas, to those converting outside, but penetrating the active gas volume and triggering the detector. A retrained set of networks with a more realistic error distribution delivers good event measurements and error estimates for our (near-) optimal weighting scheme. We test this analysis against realistic astrophysical spectra, as might be expected in early IXPE observations. 

\section{Polarization Estimation}
For clarity in the rest of the paper, we briefly summarize the steps required to go from an observed dataset of photoelectron track events measured by a detector like IXPE to the final source polarization parameters $(p_0,\theta_0)$ and their uncertainties. For additional details and proofs, one should read \citet{kislat_analyzing_2015} and P21.

\begin{figure*}
    \centering
    \includegraphics[width=0.9\textwidth]{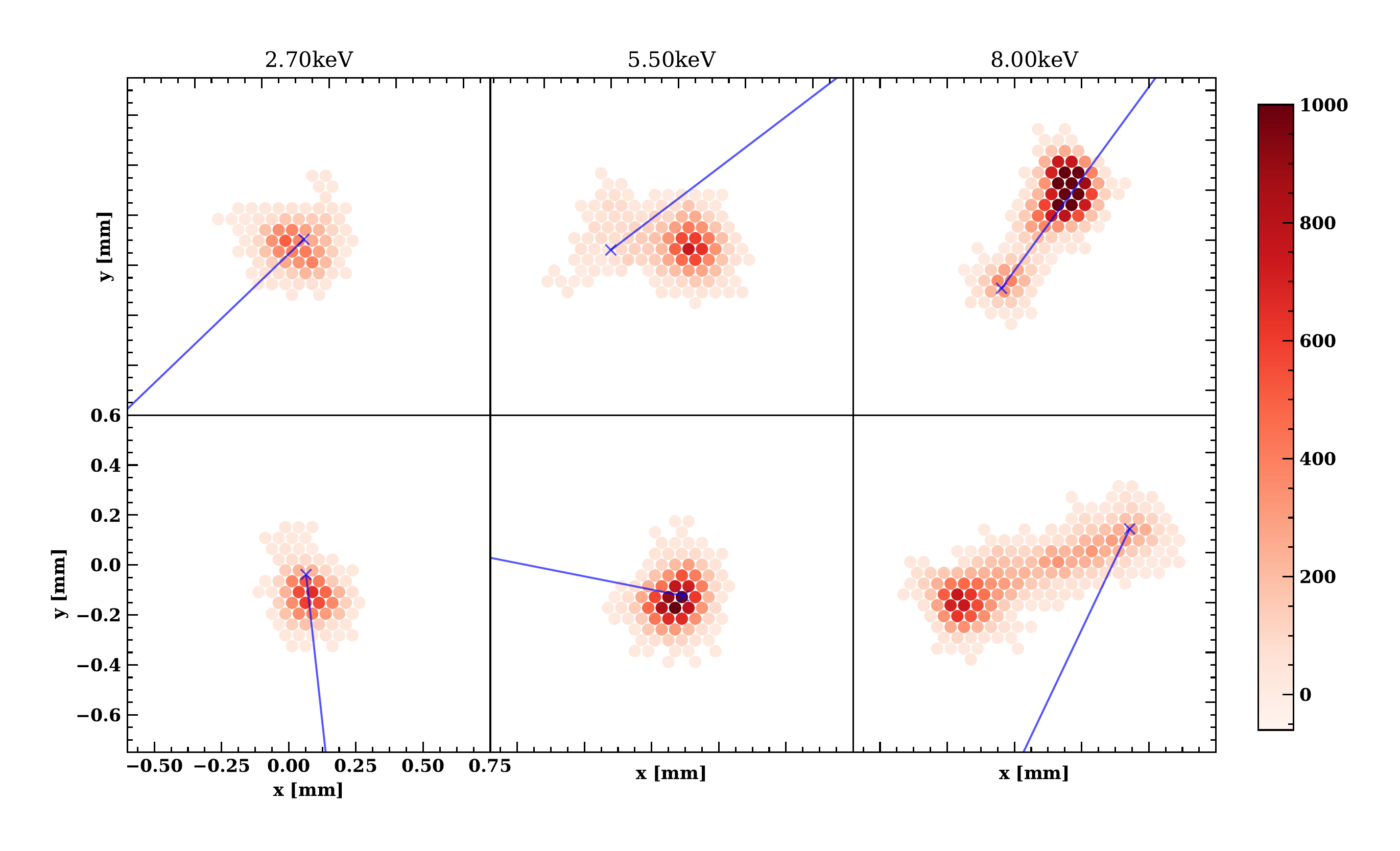}
    \caption{A selection of measured photoelectron track events from IXPE simulations. Pixel color density represents charge deposited, blue crosses show photon x-y conversion point and blue lines the initial photoelectron direction. Track morphologies vary widely and depend strongly on energy.}
    \label{fig:hex}
\end{figure*}

Imaging X-ray polarimeters measure photoelectron tracks, like those in Figure~\ref{fig:hex}. In IXPE's GPDs, photoelectron ionization tracks deposit charge onto a 2D hexagonal pixel detector array. 

From an observed set of $N$ tracks, one estimates a set of initial photoelectron directions $\{\hat{\theta}_i\}_{i=1}^N$. This can be done using, for example, the moment analysis \citep{bellazzini_novel_2003} or a NN (P21). The polarizaton fraction $p_0$ and EVPA $\theta_0$ can be reconstructed from the observations $\{\hat{\theta}_i\}_{i=1}^N$ in a number of ways, for example by binned fitting or numerical maximum likelihood estimation (MLE) for the distribution in Eq.~\ref{eqn:prob}. We prefer and recommend the approach taken by \citet{kislat_analyzing_2015}. 
They rewrite eq.~\ref{eqn:prob} as a function of linear Stokes' parameters

\begin{equation}
    \theta \sim \frac{1}{2\pi} \big(1 + \mathcal{Q}\cos2\theta + \mathcal{U}\sin2\theta \big),
    \label{eqn:prob_stokes}
\end{equation}

and they define unbiased estimators for the Stokes' parameters

\begin{equation}
    \hat{I} = \sum_{i=1}^N1 = N,
    \label{eqn:i}
\end{equation}
\begin{equation}
    \hat{\mathcal{Q}} = \frac{1}{\hat{I}}\sum_{i=1}^N2\cos2\hat{\theta}_i,
    \label{eqn:q}
\end{equation}
\begin{equation}
    \hat{\mathcal{U}} = \frac{1}{\hat{I}}\sum_{i=1}^N2\sin2\hat{\theta}_i,
    \label{eqn:u}
\end{equation}

from which the polarization fraction and EVPA estimates can be derived:

\begin{equation}
    \hat{p}_0 = \frac{1}{\mu_{100}}\sqrt{\hat{\mathcal{Q}}^2 + \hat{\mathcal{U}}^2},
    \label{eqn:p}
\end{equation}
\begin{equation}
    \hat{\theta}_0 = \frac{1}{2}\arctan\frac{\hat{\mathcal{U}}}{\hat{\mathcal{Q}}}.
    \label{eqn:th}
\end{equation}

The advantage of this approach is that we have simple analytical expressions for the desired polarization parameters and their errors. It can be shown in the high $N$ limit that $\hat{\mathcal{Q}},\, \hat{\mathcal{U}}$ are minimum variance estimators and recover the MLE result for $\hat{p}_0, \hat{\theta}_0$. Additionally, Stokes parameter estimation can be readily adjusted for event weighted estimates by modifying eqns.~\ref{eqn:i}--\ref{eqn:u} to include event weights $\{w_i\}_{i=1}^N$

\begin{equation}
    \hat{I} = \sum_{i=1}^Nw_i,
    \label{eqn:iw}
\end{equation}
\begin{equation}
    \hat{\mathcal{Q}} = \frac{1}{\hat{I}}\sum_{i=1}^N2w_i\cos2\hat{\theta}_i,
    \label{eqn:qw}
\end{equation}
\begin{equation}
    \hat{\mathcal{U}} = \frac{1}{\hat{I}}\sum_{i=1}^N2w_i\sin2\hat{\theta}_i,
    \label{eqn:uw}
\end{equation}
and introducing 
\begin{equation}
    W_2 = \sum_{i=1}^Nw_i^2.
    \label{eqn:w2}
\end{equation}

The polarization fraction and EVPA are still estimated using eqns.~\ref{eqn:p}--\ref{eqn:th} but now the effective number of events is given by

\begin{equation}
    N_{\rm eff} = \sqrt{\hat{I}^2/W_2}.
    \label{eqn:neff}
\end{equation}

\citet{kislat_analyzing_2015} derive the joint error distribution (posterior) for the polarization fraction and EVPA estimators as

\begin{equation}
\begin{aligned}
    P(\hat{p}_0,& \hat{\theta}_0| p_0,\theta_0) ={} \frac{\sqrt{N_{\rm eff}}\hat{p}_0 \mu_{100}^2}{2\pi\sigma} \times \\
    &\exp\Bigg[-\frac{\mu_{100}^2}{4\sigma^2}\Bigg\{ \hat{p}_0^2 + p_0^2 - 2\hat{p}_0 p_0\cos(2(\hat{\theta}_0 - \theta_0)) \\ & - \frac{\hat{p}_0^2p_0^2\mu_{100}^2}{2}\sin^2(2(\hat{\theta}_0 - \theta_0)) \Bigg\} \Bigg],
\end{aligned}
\label{eqn:posterior}
\end{equation}
where 
\begin{equation}
    \sigma = \sqrt{\frac{1}{N_{\rm eff}} \left( 1 - \frac{p_0^2\mu_{100}^2}{2}\right)}.
    \label{eqn:sig}
\end{equation}

Confidence intervals and the MDP$_{99}$ (eq.~\ref{eqn:MDP}, the 99\% upper limit for $p_0 = 0$) are derived from this posterior probability distribution.
In cases where $\mu_{100}$ and $p_0$ are not close to 0, the Gaussian approximation for the marginalized errors below is sufficient

\begin{equation}
    \sigma(\hat{p}_0) \approx \sqrt{\frac{2-\hat{p}_0^2\mu_{100}^2}{(N_{\rm eff}-1)\mu_{100}^2}},
    \label{eqn:sigp}
\end{equation}
\begin{equation}
    \sigma(\hat{\theta}_0) \approx \frac{1}{\hat{p}_0\mu_{100}\sqrt{2(N_{\rm eff}-1)}}.
    \label{eqn:sigth}
\end{equation}

High $\mu_{100}$ and high $N_{\rm eff}$ are both desirable to minimize the errors on recovered polarization parameters.  

In this paper we present our results in $(p_0,\theta_0)$--space, as opposed to $(\mathcal{Q},\mathcal{U})$--space. We note in some cases it may be preferable to remain in  $(\mathcal{Q},\mathcal{U})$--space since it significantly simplifies the posterior, eq.~\ref{eqn:posterior}.
On account of its simplicity and optimality, we use weighted Stokes parameter estimation throughout the remainder of this paper.

\section{Optimal Event Weights}
Photolelectron angle estimates $\hat{\theta}_i$ from track images of X-ray Polarimeters like IXPE are extremely heteroskedastic: the uncertainty on $\hat{\theta}_i$ varies widely for different events $i$. This is mainly because track morphology is a strong function of energy. Low energy tracks are short and circular making it very difficult to determine $\hat{\theta}$ while high energy tracks are long, making it easier to estimate $\hat{\theta}$. Track morphologies also vary widely at the same energy [see Figure~\ref{fig:hex}]. Thus an event-weighted estimate of polarization parameters, eqns.~\ref{eqn:p}--\ref{eqn:w2}, from an observation set $\{\hat{\theta}_i\}$ would greatly improve both multiband and single-band polarization sensitivity, since some $\hat{\theta}_i$ are much more accurate than others.

We showed in P21 that event weighting is very effective at improving sensitivity. There, we trained a deep ensemble of CNNs to predict $(\hat{\theta}_i,\hat{\sigma}_i)$ pairs where $\hat{\sigma}_i$ is the total Gaussian uncertainty on each track estimate. We used $w_i = \hat{\sigma}_i^{-\lambda}$ as event weights, where $\lambda$ is a spectrum dependent tuneable parameter. This improved MDP$_{99}$ by more than 25\% over the existing moment analysis. Although this prescription worked well, it is heuristic and by no means optimal.

Recently, \citet{marshall_multiband_2021} showed that weighting events by their expected modulation factor $\mu_{100}$ optimizes the expected signal-to-noise ratio (minimizes the MDP$_{99}$). This is illustrated in the following analysis. We define the signal-to-noise ratio (SNR)

\begin{equation}
    {\rm SNR} \propto \mu_{100}\sqrt{N_{\rm eff}}. 
\end{equation}

This is simply the inverse of the MDP$_{99}$ (without constants); an optimal weighting scheme should maximize the SNR for a fixed number of events $N$. We can expand the SNR explicitly using our weighted estimators from \textsection2, eqns.\ref{eqn:p}--\ref{eqn:neff},

\begin{equation}
    {\rm SNR} \propto \sqrt{\frac{\left(\sum_{i=1}^N2w_i\cos2\hat{\theta}_i\right)^2 + \left(\sum_{i=1}^N2w_i\sin2\hat{\theta}_i\right)^2}{\sum_{i=1}^Nw_i^2}}. 
\end{equation}

Expanding, squaring and dropping constants (which do not affect maximization) we obtain

\begin{equation}
    {\rm SNR}^2 \propto \frac{\sum_{i,j,i\neq j}w_iw_j\cos2(\hat{\theta}_i - \hat{\theta_j})}{\sum_{k}w_i^2}. 
\end{equation}

The estimators $\hat{\theta}_i$ are random variables. The true values $\theta_i$, also random variables, follow the distribution eq.~\ref{eqn:prob} with $\mu_{100} = 1$ (since they are perfectly known). Assuming the $\hat{\theta}_i$ are unbiased estimators of $\theta_i$ (true for both moment analysis and NNs) we can say

\begin{equation}
    \hat{\theta}_i = \theta_i + \epsilon_i
\end{equation}

where the measurement errors $\epsilon_i$ are independent random variables drawn from the same family of distributions with 

\begin{equation}
    \mathbb{E}[\epsilon_i] = 0, {\rm Var}[\epsilon_i] = \sigma^2_i.
\end{equation}

The specific distribution of the measurement errors $\epsilon_i$ will depend on the $\hat{\theta}_i$ estimation method; however since $\hat{\theta}_i$ are periodic, $\epsilon_i$ should follow a periodic distribution. For any $\epsilon_i$ distribution with the above properties, we can find the distribution for $\hat{\theta}_i$ as the convolution of the $\theta_i, \epsilon_i$ distributions

\begin{equation}
    \hat{\theta}_i \sim \frac{1}{2\pi} \big(1 + \mu_ip_0\cos[2(\hat{\theta}_i - \theta_0)] \big),
    \label{eqn:hat}
\end{equation}

where $0 \leq \mu_i < 1$ and $\mu_i(\sigma^2_i)$.
In other words, the distribution of estimators $\hat{\theta}_i$ are the same as the distribution of the true values $\theta_i$ but with a reduced modulation factor $\mu_i$. The measurement noise will blur the sinusoidal modulation signal by a factor $\mu_i$ for the specific event $i$. 

Knowing the distributions of $\hat{\theta}_i$, eq.~\ref{eqn:hat}, we take the expectation over SNR$^2$ (dropping constant $p_0$)

\begin{equation}
    \mathbb{E}\left[{\rm SNR}^2\right] \propto \frac{\sum_{i,j,i\neq j}w_iw_j\mu_i\mu_j}{\sum_{k}w_k^2}.
\end{equation}

Finally maximizing this expression with respect to $\{w_i\}$ in the large $N$ limit we find 

\begin{equation}
    w_i = \mu_i, 
    \label{eqn:weight}
\end{equation}

i.e. the optimal weight for an event with observation $\hat{\theta}_i$ is given by its expected $\mu_{100_i}$. Note for $N \sim 20$ this already holds with high accuracy; useful polarization measurements typically have $N > 1000$.

Eq.~\ref{eqn:weight} makes intuitive sense, with each track weighted by its expected signal. \citet{marshall_multiband_2021} adopted $\mu_{100}(E)$ for the weight. This weight can estimated from simulated or calibration data with 100\% polarized photons at a range of energies.  However each energy bin requires many events for an accurate $\mu$ estimate. Further, this weight represents only the mean response for events in a given energy bin, ignoring differences in the individual event quality, and so cannot provide optimal sensitivity. In practice, analysis of simulated data shows that this approach delivers only 6-7\% improvement in MDP$_{99}$ over an unweighted analysis. In P21 an individual event-weight by a heuristic power law $w_i = \hat{\sigma}_i^{-\lambda}$ already delivers $\sim 25$\% improvement in MDP$_{99}$ over unweighted analysis; we expect optimal weights, reflecting the individual event $\mu$'s to deliver even better sensitivity.


\subsection{NN optimal event weights}

Happily, we can train an ensemble of NNs to output individual event uncertainties, which can be converted to the optimal weights $w_i = \mu_i$ if we know the distributions of the $\theta_i$ estimation errors $\epsilon_i$. P21 assumed a simple Gaussian error distribution. This is inadequate when $\epsilon_i$ is large (poorly measured tracks). Here, we take $\epsilon_i$ to follow a von-Mises distribution ${\rm VM}(0,\kappa_i)$, the Gaussian on a periodic interval,

\begin{equation}
\epsilon_i \sim \frac{1}{2\pi I_0(\kappa_i)}\exp(\kappa_i\cos2\epsilon_i).
\label{eqn:vm}
\end{equation}

Note that for large concentration $\kappa_i$ (well-measured tracks) this reverts to a simple Gaussian with $\sigma_i^2 \approx 1/\kappa$. 

We accordingly adjust our NN $\theta_i$ loss function, minimizing the negative log-likelihood of the von-Mises distribution to predict $\hat{\theta}_i, \hat{\kappa}_i$ pairs instead of the Gaussian uncertainty predictions $\hat{\theta}_i, \hat{\sigma}_i$ used in P21. The loss on input track image $\mathbf{x}_i$ with true direction $\theta_i$ is now

\begin{equation}
\label{eqn:DE_loss}
    L(\theta_i\mid\mathbf{x}_i) = {\rm log}\left[I_0(\hat{\kappa}^a(\mathbf{x}_i))\right] - \hat{\kappa}^a(\mathbf{x}_i) \cos[2(\theta_i - \hat{\theta}(\mathbf{x}_i))].
\end{equation}

The predicted $\hat{\kappa}^a_i$ are uncertainty parameters representing aleatoric uncertainty. Using a deep ensemble of NNs \citep{lakshminarayanan_simple_2017} we can additionally estimate the epistemic uncertainty arising from imperfect model specification. With the epistemic uncertainties assumed to follow ${\rm VM}(0,\kappa^e_i)$; $\kappa^e_i$ can be estimated from the output of a deep ensemble with M NNs $\{\hat{\theta}_{ij}\}^M_{j=1}$:

\begin{equation}
    \bar{R}^2_i = \left(\frac{1}{N}\sum_{j=1}^M\cos2\hat{\theta}_{ij}\right)^2 + \left(\frac{1}{N}\sum_{j=1}^M\sin2\hat{\theta}_{ij}\right)^2 
\end{equation}
\begin{equation}
    \label{eqn:epis}
    \frac{I_1(\hat{\kappa}_i^e)}{I_0(\hat{\kappa}_i^e) } = \bar{R}_i,    
\end{equation}

with the Bessel functions $I_0$ and $I_1$. The ${\rm VM}$ assumption may not be perfect. In that case a more realistic epistemic error distribution can be recovered by bootstrap analysis of the ensemble predictions (P21); using this more realistic distribution can alter the weights and improve the MLE, but in practice the gains are small.  Equation~\ref{eqn:epis} gives an implicit prescription for the maximum likelihood estimator for $\kappa_i^e$. We can convert to an equivalent Gaussian dispersion using the circular standard deviation

\begin{equation}
    \label{eqn:circ_std}
    \sigma = \sqrt{-2\log(\bar{R})}.    
\end{equation}

Combining 
$\sigma(\hat{\kappa}^a_i)$ and $\sigma(\hat{\kappa}^e_i)$ in quadrature yields the total $\hat{\kappa}_i$. The other loss function terms for track energies and absorption points are as in P21.

To convert the output $\hat{\kappa}_i$ into optimal event weights, we need the distribution of $\hat{\theta}_i = \theta_i + \epsilon_i$. Convolving the distribution of $\theta_i$ (eq.~\ref{eqn:prob} with $\mu_{100} = 1$) with $\epsilon_i$ (eq.~\ref{eqn:vm}) we find

\begin{equation}
    \hat{\theta}_i \sim \frac{1}{2\pi} \big(1 + \frac{I_1(\kappa_i)}{I_0(\kappa_i)}p_0\cos[2(\hat{\theta}_i - \theta_0)] \big).
    \label{eqn:vm_hat}
\end{equation}

Comparing to eq.~\ref{eqn:hat}

\begin{equation}
    w_i = \mu_i = \frac{I_1(\hat{\kappa}_i)}{I_0(\hat{\kappa_i})} .
    \label{eqn:weight_func}
\end{equation}

Thus, eq.~\ref{eqn:weight_func} can be used to transform our uncertainty parameters $\hat{\kappa}_i$ into optimal event weights.

For a real detector, several effects mean that these weights will not be fully `optimal'. In detail, the error distribution is unlikely to follow a simple von-Mises shape. In particular, for the GPD we find that events fall into several classes, with differing $\epsilon_i$. Further with finite sized training data sets and limited training time, our estimates of the $\kappa_i$ will be imperfect. Nevertheless, we will show the approach achieves state-of-the-art results while significantly simplifying previous NN analysis.

\begin{figure}
    \centering
    \includegraphics[width=0.47\textwidth]{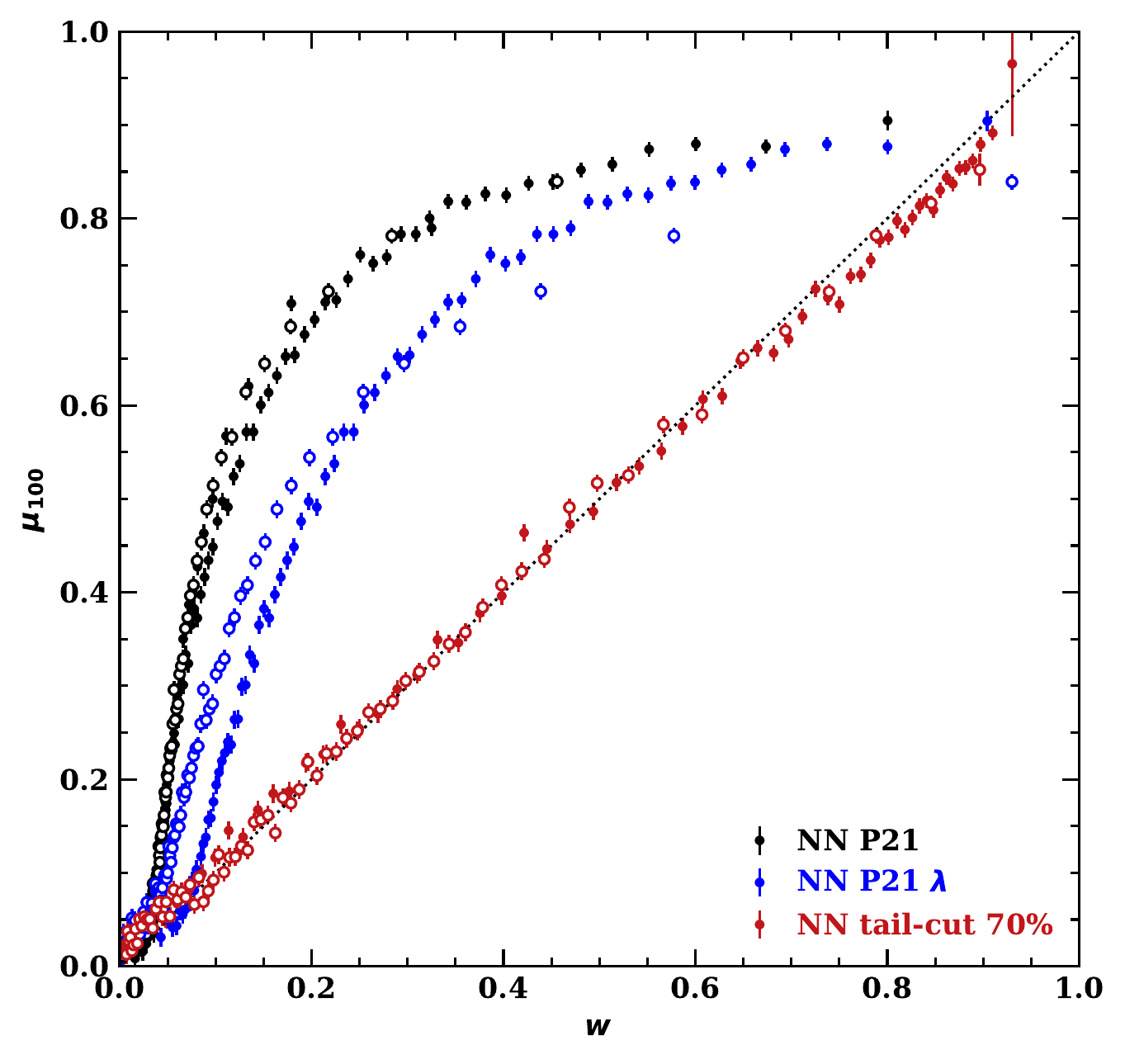}
    \caption{Measured $\mu_{100}$ as a function of weight $w$ for large test data sets of $1-10$keV simulated events. Each $\mu_{100}$ bin contains 20,000 individual track events. Open circles represent a PL1 source convolved with IXPE's effective area, closed circles a flat spectrum. For previous NN net analysis with untuned $\lambda=1$ weight (black) and with heuristic weight (blue), the weighting scheme is far from optimal and substantially dependant on the underlying spectrum. The new scheme of this paper (red) is much closer to optimal with minimal spectral dependence. }
    \label{fig:mu100}
\end{figure}

Truly optimal weights have 
$w_i = \mu_i$ for any spectrum of events. However the imperfections noted above will break this equivalence. These effects are often energy-dependent, as illustrated in Figure~\ref{fig:mu100} which shows the measured $\mu_{100}$ as a function of weight computed from (eq.~\ref{eqn:weight_func}) for two different event spectra. The spectrum-independence of our computation is imperfect, although appreciably better than for earlier schemes, such as the $\hat{\sigma}_i^{-\lambda}$ scheme or the event ellipticity weighting employed for the moment analysis.

Note that we can {\it post facto} introduce a rescaling of the $w_i$ to enforce linearity for any given source spectrum. This does indeed slightly improve MLE performance, at the expense of degraded S/N performance for other spectra. Thus, while it is acceptable to tune the $w_i$ for a typical astronomical spectrum, we do not do that here as it is better to understand and model the $\epsilon_i$ to make the analysis as close to optimal as possible, using the $w$/$\mu_{100}$ linearity as a check.

\section{Modulation Factor Recovery}

The spectral dependence in Figure~\ref{fig:mu100} emphasizes that we must normalize our polarization sensitivity across a spectrum, through $\mu_{100}$, to recover true polarization amplitudes (eq.~\ref{eqn:p}). 
Since the sensitivity varies primarily with energy we usually compute $\mu_{100} (E)$; this is complicated by the fact that the energy measured for each event $\hat{E}_i$ is an imperfect reconstruction of the true energy $E_i$. For any arbitrary spectrum with $N$ events and event weights $w_i$

\begin{equation}
    \mu_{100}= \frac{\sum^N_{i=0} w_i\mu_{100}(\hat{E}_i)}{\sum^N_{i=0} w_i} = \frac{\sum^N_{i=0} w_i^2}{\sum^N_{i=0} w_i}
\label{eqn:mu100_w}
\end{equation}

where the second equality holds when we have ideal values for the weights, avoiding all energy-dependent errors.

\begin{figure*}[t!]
    \centering
    \includegraphics[width=0.9\textwidth]{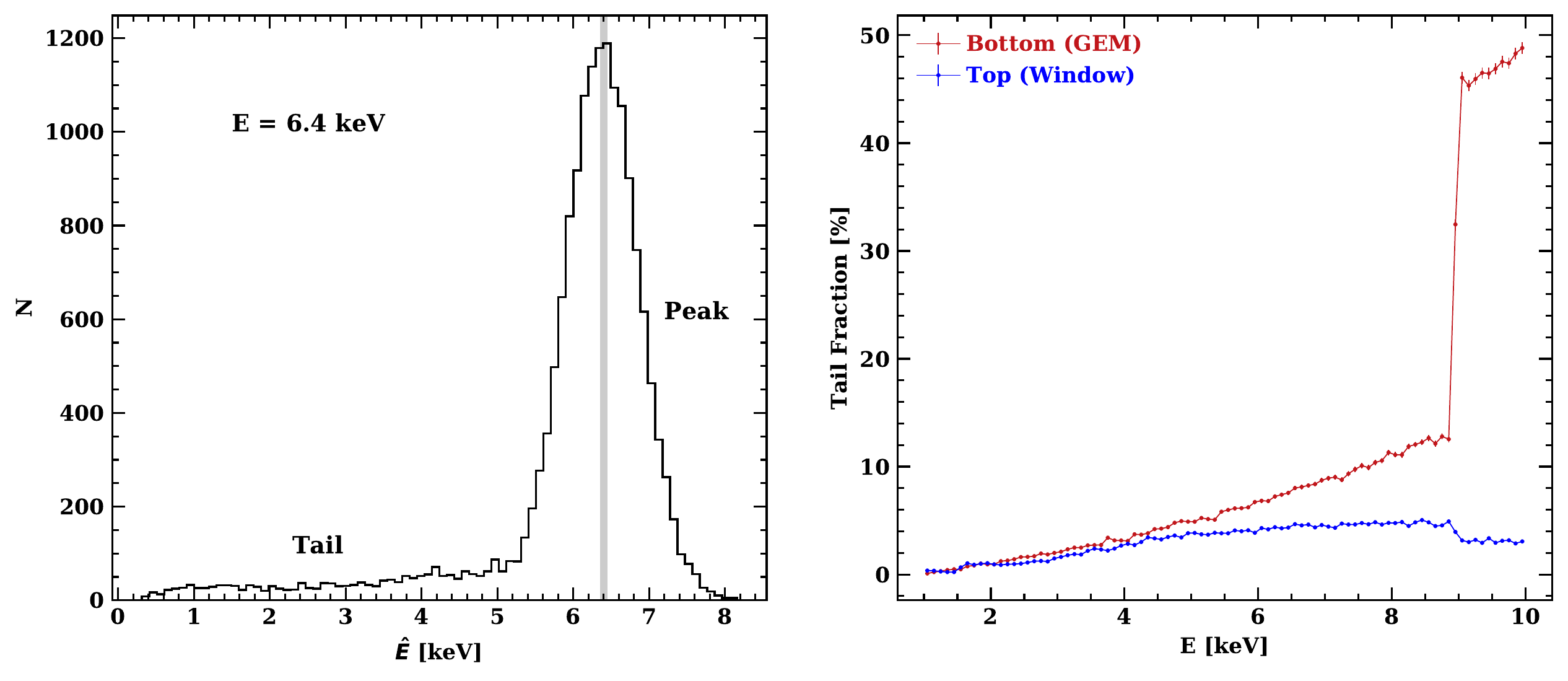}
    \caption{\textit{Left:} Recovered energy histogram for a 6.4keV line source. A simple linear function of total charge deposited is used here for recovered energy, as in the standard moment analysis. The long low energy tail is produced by events converting in the window or GEM. \textit{Right:} Fraction of events that are `tails' as a function of energy. Red and blue traces show the Be window and GEM conversion respectively. The jump in GEM conversions at 8.9\,keV is prominent.}
    \label{fig:tails}
\end{figure*}

Alas, the basic NN trained on a collection of simulated events retains substantial energy dependence. In a large part this is due to multiple event classes: photon conversions occur in the Be window at the top of the gas cell as well as on the GEM at the bottom, with electrons emerging into the gas cell, triggering the detector and generating a track. Scattering near the conversion point in the solid ensures that these tracks have little to no correlation with the initial photo-electron direction. With decreased energy deposition in the gas, these events form a `tail' to the lower end of the energy PSF (Figure~\ref{fig:tails}). These `tail' tracks represent an increasing fraction of all events at high energy with a large increase at 8.9\,keV representing the Cu absorption edge.

These `tail' events complicate spectral analysis, since a low recovered energy (low summed pixel count) event can be either a true low energy photon or the window/GEM conversion of a high energy event. In P21 the NN recognized the `tail' events as poor reconstructions with little angle and energy information, assigning such events to energies in the middle of the range to minimize recovered energy dispersion. While this did decrease the energy RMS error, as specified by the loss function, it produced a tail to the high side of low energy events, as may be seen in the black histograms on the left two panels of Figure \ref{fig:ecut}. Since most astrophysical spectra are expected to be power-laws falling with energy, this high side tail from low energy photons can significantly pollute the faint high energy bins with many low-weight poorly measured low count events assigned a mid-range energy.

\begin{figure*}[t!]
    \centering
    \includegraphics[width=0.9\textwidth]{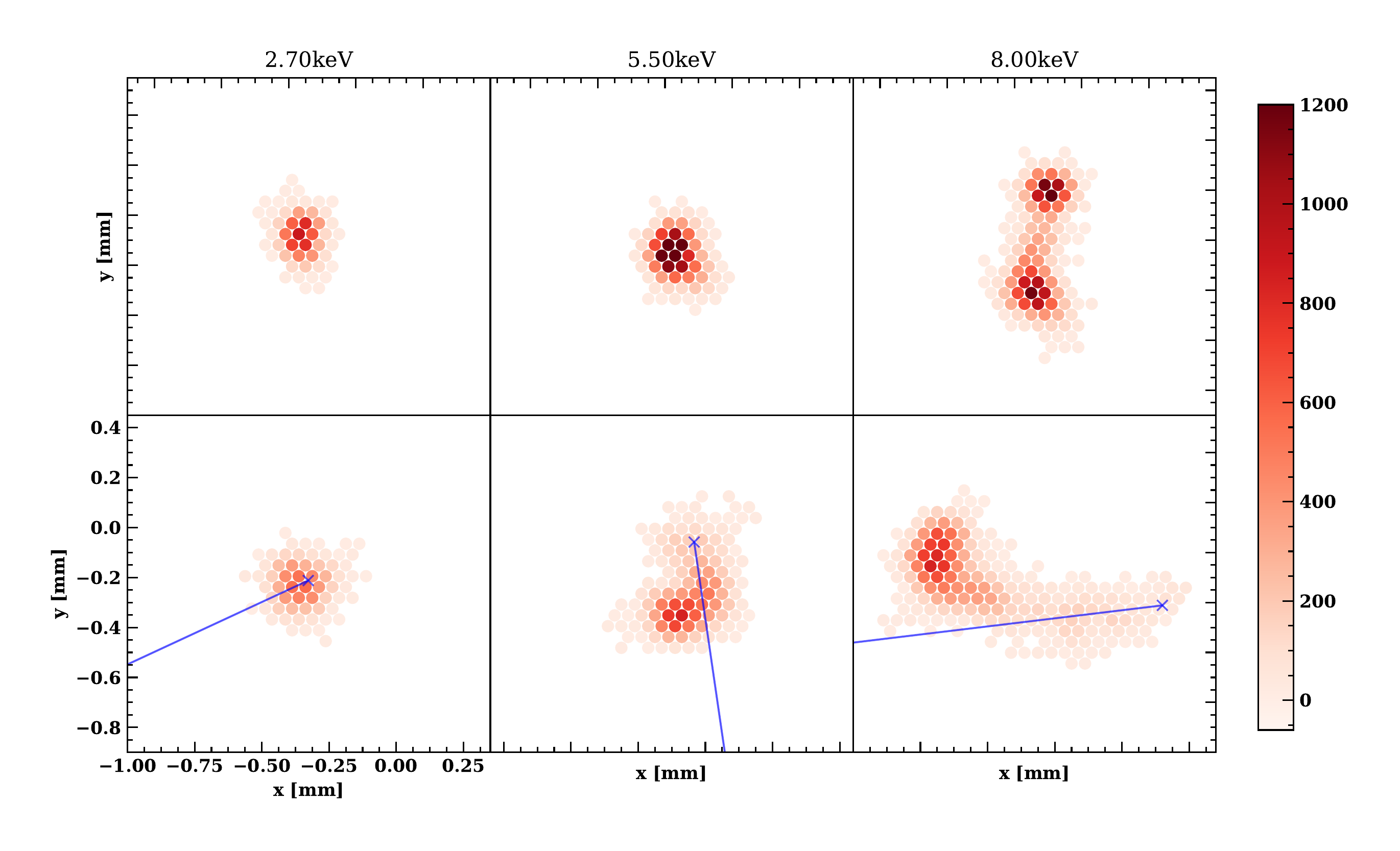}
    \caption{Example tail events (top row) and peak events (bottom row) for three different recovered energies. All plots follow the same spatial and color scale. These tail events convert in the GEM. Color denotes charge deposited in a given detector pixel. Note that tail events are more compact, short and with high charge density at maximum.}
    \label{fig:tails_v_heads}
\end{figure*}

The tail events have dramatically lower polarization sensitivity. For example, while a flat spectrum of gas events produces a (weighted) $\mu_{100} = 0.70$, the GEM conversion events have $\mu_{100} < 0.06$. Thus, when analyzed together, tail events produce broad wings in the $\epsilon_i$ distribution, departing from the \textcolor{red} peak tracks' von Mises dispersion and resulting in corrupted $w_i$ estimates. The significant energy dependence of our weights' departure from $\mu_{100}$ can be traced to the energy spectrum of tail events. In principle, with a very large training set a NN should be able to recognize different event classes, adjust the error distributions to match their varying uncertainties and deliver robust $\mu$ estimates. Limited training data and training time make this impractical.

\begin{figure*}[t]
    \centering
    \includegraphics[width=0.95\textwidth]{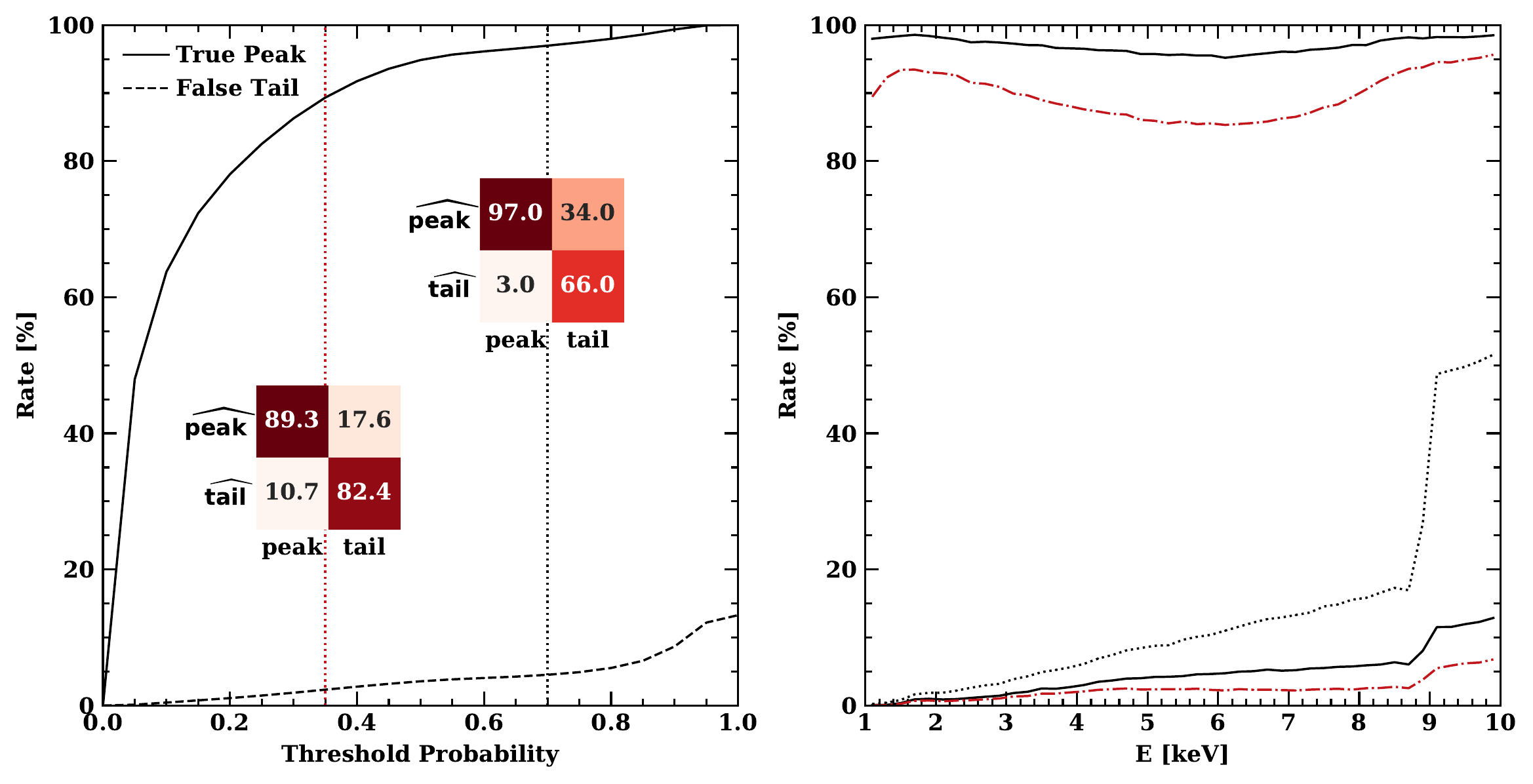}
    \caption{\textit{Left:} The solid curve shows the fraction of peak events retained as a function of the tail probability cut, while the dashed curve shows the fraction of the cut sample from the remaining tail events. Insets show the confusion matrices, normalized by column, for our adopted 70\% cut and a 35\% cut. \textit{Right:} The top curves show the peak retention as a function of energy (black solid -- 70\% cut, red dot-dash -- 35\% cut). Below, the dotted curve shows the uncut fraction of the sample due to tail events, while the lower black and red curves show the residual tail pollution (70\% and 35\% cut, respectively). Depending on how harmful tail events are to the desired measurements, different cut levels could be appropriate.}
    \label{fig:conf}
\end{figure*}

\begin{figure*}[t]
    \centering
    \includegraphics[width=1\textwidth]{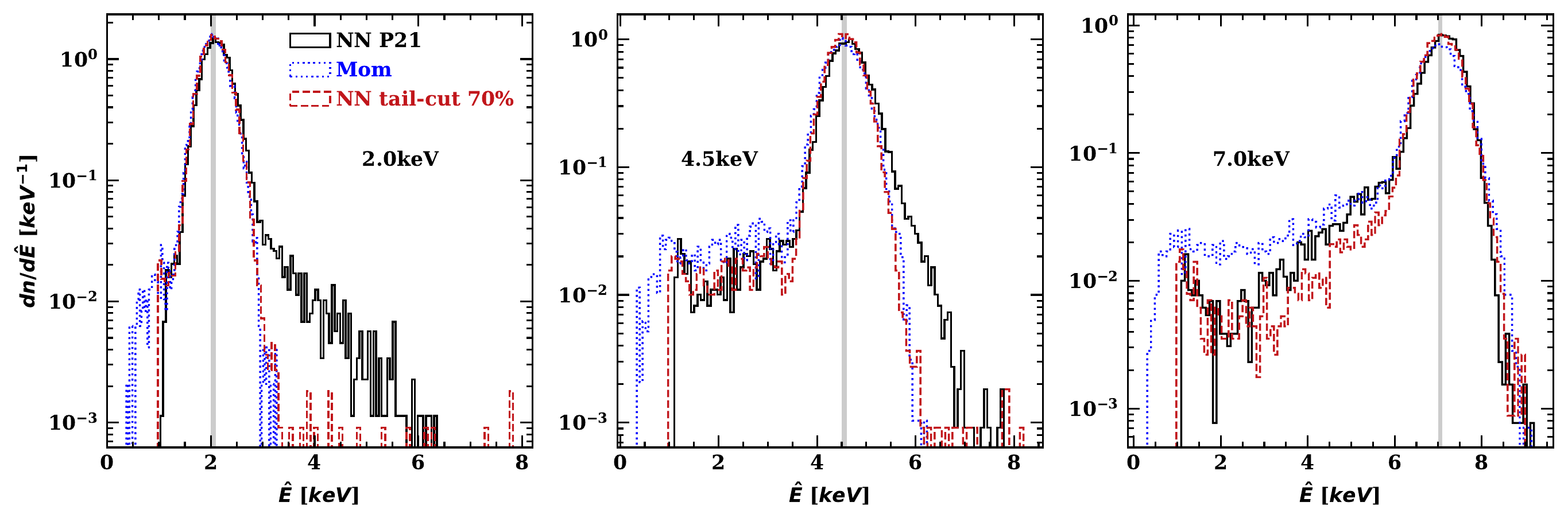}
    \caption{Response for three true energies. Note that while the P21 analysis suppressed the large tail rate seen in the Moments processing, events leaked to a high energy tail for medium to low true energies. Our morphological tail cut further suppresses the GEM/window tail, avoids the high energy leakage and achieves comparable or better peak width.}
    \label{fig:ecut}
\end{figure*}

The solution is to attempt to excise such events. The tail event tracks differ in morphology from those peak events converting in the gas (Figure~\ref{fig:tails_v_heads}). This figure compares events with the same recovered energy (largely determined through the summed pixel counts, the summed energy deposition in the gas). Tail events are typically due to photo-electrons ejected close to the GEM/window normal. Their tracks are thus more compact, with higher counts/pixel. Window events have larger drift diffusion than Cu GEM conversions. A NN can be trained on simulated events to recognize these differences. We have thus trained a `pre-processing' CNN to do such basic event classification, assigning a tail probability for each event. This CNN shares a similar ResNet architecture \citep{he_deep_2015} as the deep ensemble and is trained using a binary cross-entropy loss function. The output is a single scalar between 0 and 1 that represents the predicted event probability of peak or tail.
Figure~\ref{fig:conf} shows how the fidelity of peak/tail separation varies with tail probability. GEM event identification is especially good, the remnant tail events are mostly window events blurred by diffusion. In the present analysis we adopt a simple cut at 70\% tail probability (right dotted line) and the inset gives the numerical percentages for peak and tail IDs, as well as the false-positive percentages.  One might use this morphological classification index in a more sophisticated weighting scheme, for small additional gain.

\section{Performance of `Tail'-Suppressing NN Analysis}

\begin{figure*}[t]
    \centering
    \includegraphics[width=0.9\textwidth]{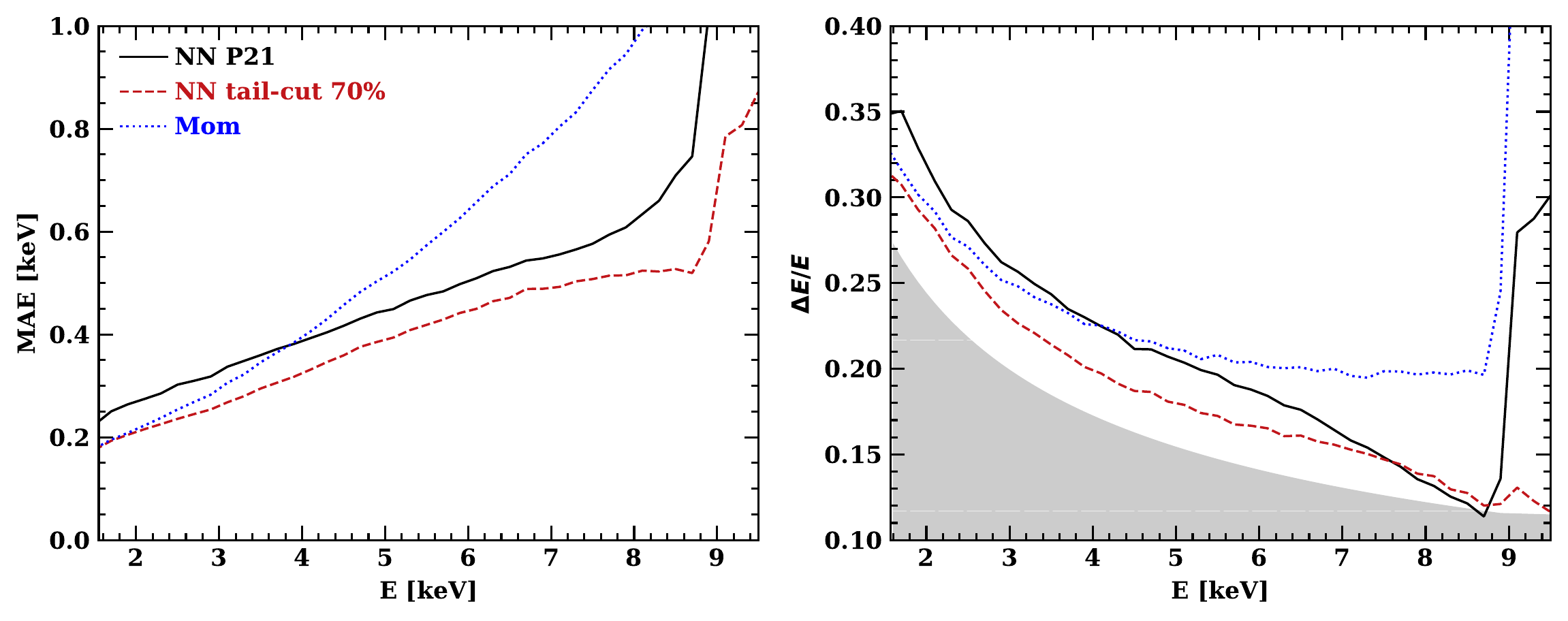}
    \caption{\textit{Left:} Mean absolute error in predicted energy as a function of true energy. \textit{Right:} FWHM (3.46$\times$ Median Absolute Deviation) of predicted energy distribution for a source at energy E. The grey band marks the limiting energy resolution in the purely-exponential multiplication regime. NNs with 70\% tail cut performs better on every metric, more accurate per track and tighter resolution at all energies. All methods suffer from the large increase of tail events above 9\,keV.}
    \label{fig:mad}
\end{figure*}

After suppression of the likely `tail' events, we have an event distribution more nearly dominated by gas/peak events. Thus we train our event measurements NNs with a set of pure gas-conversion `peak' events and analyze post-suppression data sets as if they contain no tail events. Now, training with the same loss function, we achieve improved characterization of the remaining events. Figure~\ref{fig:ecut} shows that the low-energy tail is suppressed by the cut especially at high photon energies, while the undesirable high energy tail induced at low energies by training against the un-suppressed event set is avoided. The revised NN analysis slightly improves the energy resolution in the main peak, as well. This is better displayed in Figure~\ref{fig:mad}, where two measures of the main peak width are plotted, showing that the filtered NN analysis significantly improves recovery of the true event energy. While the IXPE spectral resolution will hardly be competitive with, e.g. CCD detector spectra of these bright sources, the improved resolution does aid in measuring polarization features associated with different spectral components, as illustrated below. 

As expected, this means that the 
weight performance is also improved. Figure~\ref{fig:mu100} shows that $\mu_{100}(w)$ (red points) more closely approach the optimal weighting behavior. Following P21 we can use MDP$_{99}$ as a measure of the sensitivity. Table 1 illustrates, for two simple power-law spectra passed through IXPE's energy response, how the filtered, $\mu$-weighted NN analysis improves over both classic moments analysis and the P21 $\hat{\sigma}_i^{-\lambda}$ weights.

Note that the tail-suppressed
spectral results are for a simple 70\% tail probability cut on the event classification. Lower cuts of course improve the purity of peak identification at the cost of total event number (i.e. system effective area). 
With our present imperfect tail identification, we find that the peak events excluded by tail cuts rapidly increase MDP$_{99}$. In fact for PL2 we find a minimum MDP$_{99}=3.73\%$ with a 95\% confidence tail cut. This is $<0.6$\% (relative) MDP$_{99}$ improvement and the cut is too mild to improve the energy PSF. Accordingly, we adopt a 70\% cut to significantly improve the spectral results (cf. Figure \ref{fig:ecut}) with minimal $2$\% (relative) MDP$_{99}$ loss. An experiment seeking only broad-band polarization might impose weaker cuts. Conversely, as we improve the tail cut fidelity, stronger cuts will enhance spectral performance with smaller MDP loss. Eventually weighing with the classification NN result might be preferred, but combining spectral purity with polarization performance leads to a global performance metric whose spectrum-dependant value and error dispersion are not easily characterized. Thus an `optimal' weight including pre-filtering factors is difficult to construct. However we find that our present MDP$_{99}$ performance is good and close to minimal for a range of spectral indices, so our construction is 
near-optimal for our present classification accuracy and $\kappa$ resolution.

\begin{table}[b!]
\centering
\begin{tabular}{@{}l l l @{}}
\toprule
{\textbf{Spectrum}}&{\textbf{Method}}&MDP$_{99}$(\%) \\
\midrule
{\textbf{PL2}} & { Mom. w/ Ellip. weights} & {4.61 $\pm$ 0.02}\\ 
 & { NN w/ P21 $\lambda$ weights} &{4.09 $\pm$ 0.02} \\ 
& { NN w/ opt. wts.} &{3.75 $\pm$ 0.02 $\leftarrow$}\\
& { NN w/ opt. wts. 70\%} &{3.83 $\pm$ 0.02}\\
\midrule
{\textbf{PL1}} & { Mom. w/ Ellip. weights} & {4.15 $\pm$ 0.02}\\ 
& { NN w/ P21 $\lambda$ weights} &{3.65 $\pm$ 0.02} \\ 
& { NN w/ opt. wts.} &{3.38 $\pm$ 0.01 $\leftarrow$}\\
& { NN w/ opt. wts. 70\%} &{3.44 $\pm$ 0.01 }\\
 \bottomrule
\end{tabular}
\caption{Sensitivity analysis for two power law spectra ($dN/dE \sim E^{-N}$; PL2 for $N=2$, PL1 for $N=1$) each normalized to produce $10^5$ 2-8\,keV photons when folded through IXPE's energy response. ${\rm MDP}_{99}$ gives the sensitivities for the various weights and cuts; smaller MDP$_{99}$ is better. Mom. denotes moments analysis weighted by event ellipticity. NN denotes neural net analysis; $\lambda$ is the approximate weight defined in P21.} 

\label{tab:fom2}
\end{table}

\section{Astrophysical Spectra}

Most astrophysical spectra are moderate index power laws. When observing faint sources with a single emission process across the IXPE band (simple power-law), optimal weighting analysis is especially valuable at combining signals from a wide energy range and a wide variety of weights to derive polarization detections or upper limits on weakly polarized sources. To treat these spectra we employ $\mu_{100}({\hat E})$, computed for a PL1 spectrum, as in Figure 2. The traditional measure of polarization performance is the 99\% upper limit of a $f_{X(2-8keV)} = 10^{-11} {\rm erg\,cm^{-2}\,s^{-1}}$ power law source. For connection with earlier results we present here in Table 1 the performance for $10^5$ photons collected for two power law photon indices. We see that the heuristic P21 $\lambda$ weighting provided good sensitivity improvements over the traditional Moments method (even with ellipticity-derived weights). However, weight imperfection in both techniques restricted analysis to events with $E>2$\,keV. Our near-optimal weighting scheme allows us to extend the analysis to the full 1-10\,keV range while improving the weights themselves; this provides additional MDP$_{99}$ decrease, with about half from the increased energy range. Cutting the worst of the tail events can provide a small additional sensitivity boost. In practice, with our present event classification, dropping even the worst tail events (a 95\% cut) provides only a very small $<1$\% MDP$_{99}$ decrease. Instead, we find that improvements in energy resolution and spectrum recovery require substantially stronger tail exclusion. Thus we adopt a 70\% tail probability cut, which affords good spectral performance, but does entail a small MDP$_{99}$ increase.

IXPE will also observe sources that are very soft and very hard or have strong cut-off and line features. Although the most sensitive analysis will always involve forward modeling polarized spectra through the IXPE response and constraining model parameters via fits in data space, our revised weighting scheme is sufficiently robust to the varying event energy distributions to allow good analysis of a range of astrophysical spectra. We illustrate with a few examples, representative of very bright sources with extreme spectral parameters that are reasonably handled by our analysis.

\subsection{Soft, broken spectra -- ISP Polarization}

\begin{figure*}[t]
    \centering
    \includegraphics[width=1\textwidth]{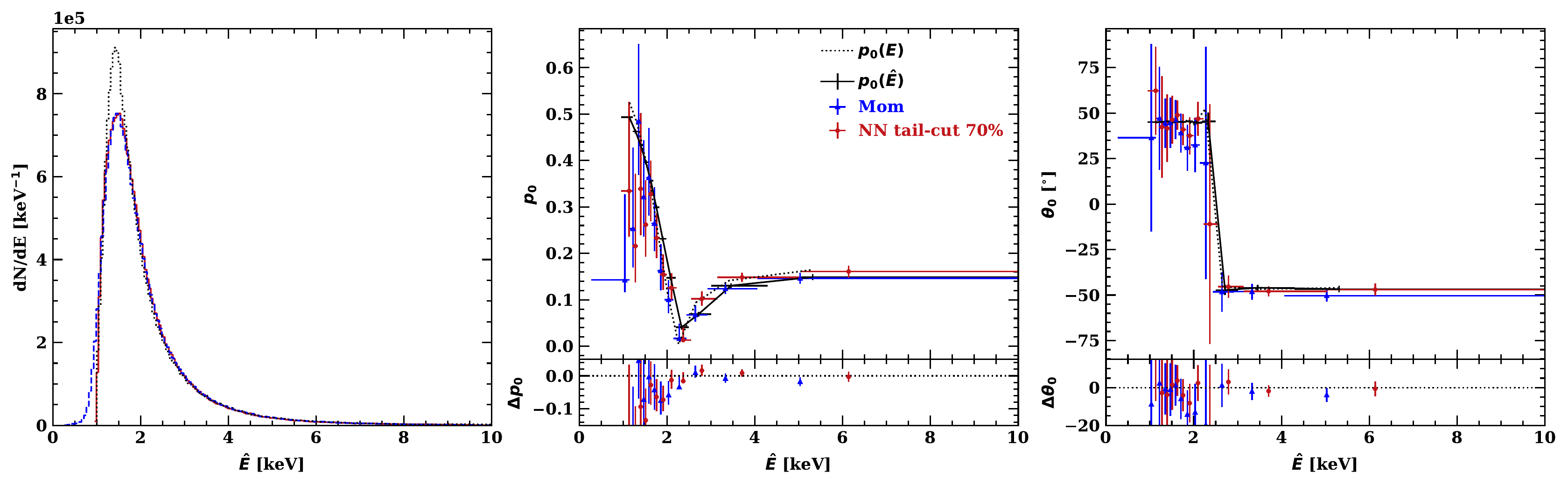}
    \caption{Simulation of an ISP with an exponential cut-off synchrotron spectrum passing to a weakly polarized Compton continuum at high energy. The dotted lines show the count spectrum, $p_0$ and PA, folded through the IXPE effective area plotted as a function of the true photon energy. The solid lines in each panel show quantities plotted as a function of the recovered photon energy, the count spectra and polarization properties suffer blurring from the methods' energy redistribution function. The residual plots at bottom show the smaller errors and modestly better recovery of the original polarization.}
    \label{fig:isp}
\end{figure*}

One source class of particular interest to IXPE are the Intermediate Sychrotron Peak (ISP) blazars. Blazars are jet-dominated AGN with spectral energy distribution (SED) dominated by synchrotron emission at low energies and Compton emission at high energies. For ISPs the transition occurs at soft X-ray energies in the IXPE band. With an appropriate cut-off energy, the synchrotron component from a very few zones will dominate in the low energy IXPE band. This exponentially cut-off synchrotron emission can be very steep and very highly polarized \citep{peirson_polarization_2019}.  At higher energies, the inverse Compton emission will have a faint hard spectrum and a low $p_0$ at a different PA. In Figure \ref{fig:isp}, we show a model whose highest energy synchrotron component cuts off at 0.25\,keV with Compton emission at an orthogonal PA dominating above $\sim 3$\,keV. We contrast the polarization recovery of the standard moments method and our new optimal weight analysis. The total number of counts generated for the spectrum between 1-10keV is equivalent to an IXPE 10-day exposure of a $10^{-10}$ erg/cm$^2$/s 2-8keV source.

For these very bright models 
IXPE should be able to make good polarization measurements independent of analysis method, and both moments analysis and optimal weighting are expected to give interesting polarization results. Although the optimal weighting scheme shows to greatest advantage when acting on a broad spectral band with a range of weights (as in Table 1), we do see see significant gains in some aspects. First the PA/$p_0$ error bars are, as expected, 20-30\% smaller than for Moments analysis. The optimal weighting with cuts also minimizes the artificial low energy tail to the moments polarization signal. The improved energy resolution of optimal weight analysis also better resolves the PA switch near 2\,keV.

\subsection{Hard, absorbed spectra -- Bright Accretion Powered Binaries}

Bright X-ray binaries are also interesting targets for IXPE. Here we simulate a source whose spectrum resembles the pulse-average spectrum of the bright $f_{2-10keV} \approx 10^{-9} {\rm erg\, cm^{-2}s^{-1}}$ HMXB (High Mass X-ray Binary) GX 301$-$2. We take spectral parameters from \citet{pravdo_asca_1995}; the absorption in the accretion flow is large, leading to a hard spectrum, cut off sharply below 3\,keV and we ignore the observed faint low energy emission. The spectrum also features a strong Fe K$\alpha$ line at 6.41\,keV and the absorption edge at 7.11\,keV. We assume that the continuum is strongly (1/3) polarized. The Fe K$\alpha$ line is unpolarized except for weak polarization in a Compton shoulder of scattered flux centered at 6.3\,keV (which, unfortunately, IXPE does not resolve). For this spectrum the total number of counts between 1-10keV is equivalent to an IXPE 2.5-day exposure of a $2 \times 10^{-9}$ erg/cm$^2$/s 2-8keV source.
Both moments and NN have some difficulty recovering polarization below 4\,keV. This is a result of the very large contamination of tail events at low energies, many arising from GEM conversions above
9\,keV.

\begin{figure*}[t]
    \centering
    \includegraphics[width=1\textwidth]{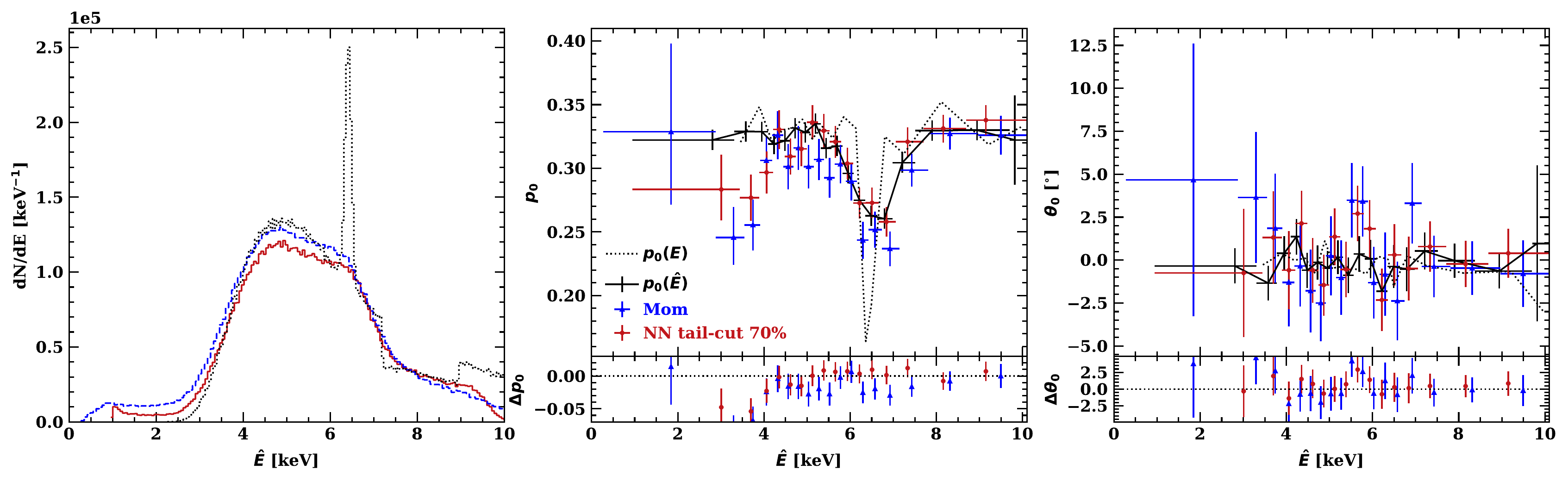}
    \caption{Simulation of a hard, absorbed HMXB similar to GX 301$-$2.  As for Figure \ref{fig:isp} the dotted lines show source properties as a function of the true photon energy, while the solid lines show the expected signal, and the various methods' recovery of this signal as a function of the method-estimate recovered energy. Our NN analysis with tail cut decreases the tail counts at low energy, and improves the isolation of the $p_0$ dip associated with the unpolarized Fe K line at 6.4\,keV (which is barely seen in the recovered count spectra). The residual plots at bottom show better recovery of PA/$p_0$ with smaller errors.}
    \label{fig:gx301}
\end{figure*}

\section{Conclusion}

We have shown that an optimal weighting of track position angles (of known error distribution) recovers the source polarization with the highest possible sensitivity and resolution-limited independence from the underlying spectral form. In practice for the IXPE GPD detector to approach this theoretical limit, we needed to extend our track analysis to separate events converting in the detector gas from the corrupted tracks of events converting in the detector walls. This, and our track geometry characterization, are accomplished by an ensemble of convolutional neural nets. Although track characterization is imperfect, we find that our analysis improves polarization sensitivity, approaching the `optimal' limit, while slightly improving energy resolution. In addition the technique allows us to extend the {\it IXPE} sensitive band obtaining useful signal from $\sim$1\,keV to $>10$\,keV.

Applying our our analysis to simulated power law spectra we show $20-25$\% sensitivity improvement over the standard (weighted Moments) technique and $\sim 8-9$\% improvement over our prior neural net analysis. The robustness of the measurements are illustrated by simulating spectra from especially soft (ISP Blazar) and hard (absorbed neutron star binary) spectra from source classes planned for early IXPE observation. The `optimal' weighting shows modestly improved recovery of the model polarized spectrum at both extremes. Of course, these results need to be validated with real flight GPD data, including all the peculiarities of the as-built detectors. But in view of the improvements relative to existing methods, our analysis of simulated GPD events show that the prospects for improved imaging polarimetry with IXPE and similar systems are good. 

\acknowledgments

We thank Luca Baldini and the anonymous referee for for careful readings of the text, Alessandro Di Marco, Fabio Muleri and Paolo Soffitta for advice on IXPE calibration, Herman Marshall for discussion of polarization statistics, and Giorgio Matt for discussion of source spectra. This work was supported in part by the NASA FINESST program (grant 80NSSC19K1407) and grant NNM17AA26C from the Marshall Space Flight Center.

\bibliography{references}
\bibliographystyle{aasjournal}
\end{document}